\documentclass [12pt,a4paper]{article}

\title{ANALYSIS OF PODKLETNOV'S PHENOMENON BASED ON EXPANSIVE
  NONDECELERATIVE UNIVERSE}
\author{Miroslav S\'{u}ken\'{\i}k and Jozef \v{S}ima \\[1ex]
  Slovak Technical University, FCHPT, Radlinsk\'{e}ho 9, \\
  812 37 Bratislava, Slovakia
  \\
  e-mail:  sukenik@minv.sk,sima@chtf.stuba.sk} \date{}
\begin{document}
\maketitle
\begin{abstract}
Owing to Vaidya metric including, the model of Expansive Nondecelerative 
Universe (ENU) is able to localize gravitational energy. Based on the 
quantification of gravitational field of the Earth, ENU allows to 
rationalize and quantify the effects of a superconductor-based high voltage 
impulse gravity generator constructed by Podkletnov. The differences in 
energy effect observed for two generators used are explained and improvement 
of the experiment arrangement is proposed.


\end{abstract}

\section{Introduction}

In the recent papers [1-3], the detection of anomalous forces formed within 
electron discharge from a superconductive ceramic emitter to a targeting 
electrode has been described. It has been hypothesed that these forces are 
of gravitational nature. The whole time of a discharge was $10^{ - 5}$ to 
$10^{ - 4}$~s, the peak value of the current at the discharge was of the 
order $10^{4}$ A, the voltage varied in the range of 500 kV to 2 MV. The 
distance between the electrodes was 15 cm to 2 m. Based on the total charge 
localized on the emitter ($\sim $ 0.1 C) the discharge energy approached 
$10^{5}$ J. The gravity impulse accompanying the discharge propagated as a 
coherent beam in the same direction as the discharge and penetrated through 
different media (air, brick wall, steel plate) without any noticeable loss 
of energy. It acted on mobile objects such as spherical 10 to 50 g weighing 
pendulums made from different materials (rubber, glass, metal, plastics) 
hanging on a cotton thread like a repulsive force independent on the 
pendulum material and proportional to their mass. Measurements of the 
impulse taken at the distance of 3 m, 6 m and 150 m gave identical results. 

The experiment was theoretically rationalized by Modanese [3] who stated 
that the phenomenon could not be understood in the framework of general 
relativity. He proposed an explanation combining a quantum gravity approach 
and anomalous vacuum fluctuations. 

In this contribution, an independent rationalization of the physical nature 
of the Podkletnov's experiments, explanation of different efficiency reached 
using two generators, and proposals for improvement of the efficiency are 
offered stemming from the background of the model of Expansive 
Nondecelerative Universe [4-8]. 

\section{Background of the model of Expansive Nondecelerative Universe}

The basic principles of the Expansive Nondecelerative Universe model were 
presented in a series of papers [4-8]. The model differs from more 
frequently used models of inflationary universe in the following features:

\noindent
a) Schwarzschild metric is replaced by Vaidya metric, 

\noindent
b) the Universe permanently expands by the velocity of light $c$, 

\noindent
c) simultaneous creation of matter and the equivalent amount of 
gravitational energy (which is, however, negative and thus the total value 
of mass-energy is constant and equal zero) occurs. 

\noindent
d) the Einstein cosmological constant and the curvature index are of zero 
value.

Statement b) can be expressed as follows
\begin{equation}
\label{eq1}
a = c.t_{c} = {\frac{{2G.m_{U}}} {{c^{2}}}}
\end{equation}
where $a$ is the gauge factor, $t_{c} $ is the cosmological time, $m_{U} $ 
is the Universe mass (their present values are $a \cong 1.3\times 10^{26}$ m, 
$m_{U} \cong 8.6 \times 10^{52}$ kg, $t_{c} \cong 1.4\times 10^{10}$ yr). 

In the ENU the state function is formulated as
\begin{equation}
\label{eq2}
p = - {\frac{{\varepsilon}} {{3}}}
\end{equation}
i.e. a trace of the energy-momentum tensor equals to zero. 

In Vaidya metric [9, 10] the line element is formulated in the form
\begin{equation}
\label{eq3}
ds^{2} = {\frac{{\Psi '^{2}}}{{f_{(m)}^{2}}} }\left( {1 - {\frac{{2\Psi 
}}{{r}}}} \right).c^{2}.dt^{2} - \left( {1 - {\frac{{2\Psi}} {{r}}}} 
\right)^{ - 1}dr^{2} - r^{2}\left( {d\theta ^{2} + \sin ^{2}\theta .d\varphi 
^{2}} \right)
\end{equation}
where
\begin{equation}
\label{eq4}
\Psi = {\frac{{G.m}}{{c^{2}}}}
\end{equation}

$m$ is the mass of a body, $f_{(m)} $ is an arbitrary function. In order to 
$f_{(m)} $ be of nonzero value, it must hold

\begin{equation}
\label{eq5}
f_{(m)} = \Psi {\left[ {{\frac{{d}}{{dr}}}\left( {1 - {\frac{{2\Psi 
}}{{r}}}} \right)} \right]} = {\frac{{2\Psi ^{2}}}{{r^{2}}}}
\end{equation}

As rationalized in [11] in the ENU

\begin{equation}
\label{eq6}
\Psi ' = {\frac{{d\Psi}} {{c.dt}}} = {\frac{{\Psi}} {{a}}}
\end{equation}

Vaidya metric may be applied in all cases for which the gravitational energy 
is localizable, i.e. if

\begin{equation}
\label{eq7}
r \le r_{ef} 
\end{equation}

\noindent
where $r_{ef} $ is the effective gravitational range of a body with the 
gravitational radius $2\Psi $.

\begin{equation}
\label{eq8}
r_{ef} = \left( {r_{g} .a} \right)^{1 / 2} = \left( {2\Psi .a} \right)^{1 / 
2}
\end{equation}

Gravitational influence can be thus realized only when the absolute value of 
gravitational energy density will exceed the critical energy density. 

The energy-momentum complex of Einstein pseudotensor is [12]

\begin{equation}
\label{eq9}
\theta _{i}^{k} = {\frac{{1}}{{16\pi}} }{\left[ {{\frac{{g_{in}}} {{\sqrt { 
- g}}} }{\left\{ { - g(g^{kn}.g^{lm} - g^{\ln} .g^{km})} \right\}},_{m}}  
\right]}_{,_{l}}  
\end{equation}

Application of Vaidya metric in Cartesian Kerr-Schild coordinates was solved 
by Virbhadra [10] who calculated the components of Einstein pseudotensor in 
a general form. In our approach, instead of a generally formulated $\Psi '$ 
its actual expression (\ref{eq6}) is offered. The complete pseudotensor components 
are published elsewhere [11], in case of weak fields, its main component

\begin{equation}
\label{eq10}
\theta _{0}^{0} = - {\frac{{c^{4}}}{{8\pi .G.r^{2}}}}\Psi '
\end{equation}

\noindent
represents the gravitational energy density in the first approximation. 
Relation (\ref{eq10}) may be understood as being identical to Tolman equation

\begin{equation}
\label{eq11}
\varepsilon _{g} = - {\frac{{R.c^{4}}}{{8\pi G}}}
\end{equation}

\noindent
where $R$ is the scalar curvature. In Schwarzschild metric, $R = 0$, $i.e$. 
gravitational energy is not localizable outside a body since $\varepsilon 
_{g} = 0$ there. In Vaidya metric

\begin{equation}
\label{eq12}
R = {\frac{{6G}}{{r^{2}.c^{3}}}}.{\frac{{dm}}{{dt}}} = {\frac{{6G.m}}{{t_{c} 
.r^{2}.c^{3}}}} = {\frac{{3r_{g}}} {{a.r^{2}}}}
\end{equation}
where $R$ is the scalar curvature in the distance$ r$ for a body
having the mass $m$. It follows from (\ref{eq6}), (\ref{eq10}) and
(\ref{eq11}) that the density of the gravitational field energy
generated by a body with the mass $m$ at the distance $r$ is
\begin{equation}
\label{eq13}
\varepsilon _{g} \cong - {\frac{{3m.c^{2}}}{{4\pi .a.r^{2}}}}
\end{equation}

Within the limits of ENU model it is thus possible to localize and determine 
$\varepsilon _{g}$.

Substituting the Earth mass and its radius into (\ref{eq13}), the
energy density value of the Earth gravitational field is then
\begin{equation}
\label{eq:14a}
{\left| {\varepsilon _{g(Earth)}}  \right|} \cong 25\mbox{J.m}^{ - 3} 
\end{equation}
Stemming from (\ref{eq11}), gravitational output $P_{g} $ is

\begin{equation}
\label{eq14}
P_{g} = - {\frac{{d}}{{dt}}}\int {{\frac{{R.c^{4}}}{{8\pi .G}}}dV = - 
{\frac{{m.c^{3}}}{{a}}}} = - {\frac{{m.c^{2}}}{{t_{c}}} }
\end{equation}

The energy density of the gravitational field can be expressed also as

\begin{equation}
\label{eq15}
\varepsilon _{g} = {\frac{{E_{g}}} {{\lambda _{C}^{3}}} }
\end{equation}

\noindent
where $E_{g} $ and $\lambda _{C} $ are the energy of a gravitational field 
quantum and its Compton wavelength

\begin{equation}
\label{eq16}
\lambda _{C} = {\frac{{\hbar}} {{m.c}}} = {\frac{{\hbar .c}}{{E_{g}}} }
\end{equation}

For the energy of a gravitational field quantum it then hold

\begin{equation}
\label{eq17}
E_{g} = \left( {\varepsilon _{g} .\hbar ^{3}.c^{3}} \right)^{1 / 4}
\end{equation}

Due to its wave nature, gravitational field may be described by a 
wavefunction $\Psi _{g} $

\begin{equation}
\label{eq18}
\Psi _{g} = \exp (i.\omega .t)
\end{equation}

\noindent
and using (\ref{eq17}) and (\ref{eq18})

\begin{equation}
\label{eq19}
\omega = \left( {{\frac{{m.c^{5}}}{{a.r^{2}.\hbar}} }} \right)^{1 / 4}
\end{equation}

Based on Schr\"{o}dinger-like equation for the gravitational waves

\begin{equation}
\label{eq20}
E_{g} .\Psi _{g} = i.\hbar {\frac{{d\Psi _{g}}} {{dt}}}
\end{equation}

\noindent
the energy of a quantum of gravitational field induced by a body with the 
mass $m$ at the distance $r$ is given as

\begin{equation}
\label{eq21}
E_{g} = - \left( {{\frac{{m.\hbar ^{3}.c^{5}}}{{a.r^{2}}}}} \right)^{1 / 4}
\end{equation}

Substituting the values for the Earth into (\ref{eq21}) it follows
\begin{equation}
\label{eq:23a}
{\left| {E_{g}}  \right|} \cong 10^{ - 19} \mbox{J}
\end{equation}

\section{Interpretation of the Podkletnov's results stemming from the ENU}

Within the Podkletnov's experiments [1-3], electrostatic field with the mean 
voltage intensity 
\begin{equation}
\label{eq:24a}
E \cong 2.5x10^{6} \mbox{V/m}
\end{equation}
was applied which corresponds to the mean energy density 
\begin{equation}
\label{eq:25a}
\varepsilon _{E} \cong 27 \mbox{J/m}^{3} 
\end{equation}
Comparing this value with that given by~(\ref{eq:14a}) it is obvious that
\begin{equation}
\label{eq22}
\varepsilon _{E} \cong {\left| {\varepsilon _{g(Earth)}}  \right|}
\end{equation}

Our explanation of the Podkletnov's phenomenon lies in a hypothesis that the 
Earth gravitational field interferes with the electrostatic field created 
within a discharge. The gravitational attractive force of the Earth results 
from the attraction of a body (pendulum) by the whole Earth. Provided that 
it is the only force exerting on a body, its direction is vertical as a 
result of the vector sum of attractive forces exerted by all the Earth 
parts. Should the Earth gravitational field be attenuated in a certain 
location (direction) the other attractive forces prevail. The observed 
deflection of pendulum thus results from ,,non-vertical`` Earth attraction. 
The pendulum deflection is, therefore, not a consequence of a repulsive 
force created within a discharge but lies in a local attractive force 
decreasing due to reducing the Earth gravitational field caused by its 
interference with electrostatic field. 

In the mathematical language the above ideas can be expressed as follows. 

The energy density of the electrostatic field is defined as

\begin{equation}
\label{eq23}
\varepsilon _{E} = {\frac{{\varepsilon _{o} .V^{2}}}{{2r_{x}^{2}}} }
\end{equation}

\noindent
where $V$ is the applied voltage and $r_{x} $ is the distance between 
electrodes. Based on (\ref{eq13}), (14), (\ref{eq22}), and (\ref{eq23}) for the conditions on the 
Earth surface it then follows that

\begin{equation}
\label{eq24}
V^{2} \cong {\frac{{3m_{(Earth)} .c^{2}.r_{x}^{2}}} {{2\pi .\varepsilon _{o} 
.r_{(Earth)}^{2} .a}}}
\end{equation}

It is obvious that the optimal distance between the electrodes is 
\begin{equation}
\label{eq:29a}
r_{x} \cong 0.2 \mbox{m}
\end{equation}
for the applied voltage of $V = 500$ kV, and 
\begin{equation}
\label{eq:30a}
r_{x} \cong 0.85 \mbox{m}
\end{equation}
for $V = 2$ MV. 

A theoretical treatment of (\ref{eq24}) leads to a conclusion that to
discharge only one electron from the emitor at the voltage of $V \cong
1$ V, it should hold
\begin{equation}
\label{eq:31a}
r_{x} \cong 10^{ - 7} \mbox{m}
\end{equation}
Such an electron would obtain the energy of 1 electronvolt, i.e.
\begin{equation}
\label{eq:32a}
E_{(e)} \cong 1.6 \times 10^{ - 19} \mbox{J}
\end{equation}
which is close to the energy of a gravitational quantum at the Earth
surface~(\ref{eq:23a}). It can be, therefore, supposed that in cases
when the requirement (\ref{eq22}) is fulfilled, each discharged
electron creates one quantum of the Earth gravitational field
irrespective to the voltage applied. It directly followed from
(\ref{eq17}) that the energy of a gravitational quantum depends on the
energy density only.  At the Podkletnov's experiment, the total charge
localized on the emitor at the voltage of 2 MV reached
\begin{equation}
\label{eq:33a}
Q \cong 0.1 \mbox{C}
\end{equation}
which corresponds to a number of electrons
\begin{equation}
\label{eq25}
n_{(e)} = {\frac{{Q}}{{e}}} \cong 10^{18}
\end{equation}
Within a discharge at 2 MV, $10^{18}$ quanta of gravitational field are 
formed, each of them bearing the energy of about $10^{ - 19}$ J. 

Podkletnov observed [3] that the total energy of a deflection depended on 
the pendulum mass. This observation is consistent with (\ref{eq4}) and (\ref{eq8}) stating 
that the higher the mass, the higher the effective gravitational radius. It 
must hold for the potential energy of a displaced pendulum

\begin{equation}
\label{eq26}
\Delta E = m_{(P)} .n_{(e)} .{\left| {E_{g}}  \right|}(kg^{ - 1})
\end{equation}

\noindent
where $m_{(P)} $ is the mass of pendulum. For $m_{(P)} = 18.5$ g (data in 
[3] are related to this pendulum mass) relation (\ref{eq26}) leads to
\begin{equation}
\label{eq:36a}
\Delta E \cong 1.8x10^{ - 3} \mbox{J}
\end{equation}
which is in good accordance with the experimental data obtained using both a 
newer equipment (emitter 2) 
\begin{equation}
\label{eq:37a}
\Delta E \cong 1.3x10^{ - 3}\mbox{J}
\end{equation}
and an older one (emitter 1), where
\begin{equation}
\label{eq:38a}
\Delta E \cong 2.3x10^{ - 3} \mbox{J}
\end{equation}

Stemming from (30) and comparing the values in (37) and (38) it is obvious 
that the actual distance between the electrodes in a newer equipment (0.15 
to 0.4 m) was far from the optimal distance 0.85 m. We believe that the 
rearrangement of a discharge chamber so as to permit to reach a higher 
distance between the electrodes will lead to a higher alteration in the 
pendulum potential energy. In addition, to preserve a high level of 
coherency when applying higher voltages, the space between two electrodes 
should be localized in an external magnetic field having the intensity 
proportional to $r_{x} $.

Relation (\ref{eq26}) can be obtained by an independent way too. A flow of the 
gravitational energy of the Earth, $\sigma _{(Earth)} $ through its surface 
unit (hereinafter, all data are related to the surface of one square meter) 
is

\begin{equation}
\label{eq27}
\sigma _{(Earth)} = {\frac{{P_{g(Earth)}}} {{4\pi .c.r_{(Earth)}}} }
\end{equation}

\noindent
where $P_{g(Earth)} $ is the gravitational output of the Earth (\ref{eq14}). Due to 
a close proximity of the emitter and the pendulum and owing to coherent 
nature of the pulses, it can be written

\begin{equation}
\label{eq28}
\sigma _{(P)} \cong n_{(e)} .{\left| {E_{g}}  \right|} = {\frac{{{\left| 
{E_{g}}  \right|}.Q}}{{e}}}
\end{equation}

\noindent
where $\sigma _{(P)} $ is the gravitational energy flow from Podkletnov 
equipment through a surface unit. In such a case it must hold for the energy 
of pendulum deflection

\begin{equation}
\label{eq29}
{\frac{{\Delta E}}{{U}}} = {\frac{{\sigma _{(P)}}} {{\sigma _{(Earth)}}} }
\end{equation}

\noindent
where $U$ is the potential energy of the pendulum with the mass $m_{(P)} $

\begin{equation}
\label{eq30}
U = {\frac{{G.m_{(Earth)} .m_{(P)}}} {{r_{(Earth)}}} }
\end{equation}

It follows from (\ref{eq27}) to (\ref{eq30}) that

\begin{equation}
\label{eq31}
\Delta E \cong \left( {{\frac{{4\pi .a.G.{\left| {E_{g}}  
\right|}}}{{e.c^{2}}}}} \right).Q.m_{(P)} 
\end{equation}

The expression in parentheses is a constant of the value close to 1 C$^{ - 
1}$ m$^{2}$ s$^{ - 2}$. This is why (\ref{eq31}) can be rewritten in a simpler form 
(and expressed in the above unit) as

\begin{equation}
\label{eq32}
\Delta E \cong Q.m_{(P)} 
\end{equation}

Based on (\ref{eq25}) and (\ref{eq26}), as well as on the close numerical values of ${\left| 
{E_{g}}  \right|}$ and $e$, relations (\ref{eq32}) and (\ref{eq26}) become thus almost 
identical. 

For a vertical deflection of the pendulum, $h$ it can be written

\begin{equation}
\label{eq33}
h \cong {\frac{{Q}}{{g}}}
\end{equation}

\noindent
where $g$ is the Earth gravitational acceleration. Once again, it is obvious 
from (\ref{eq33}) that the vertical deflection of the pendulum depends only on the 
total charge, i.e. on the voltage between the electrodes.

\section{Prospectives of application of the Podkletnov's phenomenon}

The Podkletnov's phenomenon seems to be in principle of qualitatively new 
nature. If theoretically and experimentally proved more deeply, it could 
form an advantageous standpoint to rationalize several open phenomena. In 
this part its connection to the problems of fire balls stability, the 
hydrogen atom stability, and solar corona temperature is briefly outlined 
(the treatment of the problems in more details will be published elsewhere).

\textbf{\textit{a) Stability of fire-ball-like plasmatic bodies}}

Our calculations indicate that in a case of simultaneous shortening the time 
of discharge and increasing the external magnetic field intensity at the 
voltage exceeding 500 kV, a stable plasmatic body with the radius of about 
10 - 30 cm, similar to a fire ball, might be formed. From the technical 
point of view such an experiment is realizable since the kinetic energy of 
an electron at $V \ge 500$ kV is comparable to its rest energy. The success 
of the experiment is conditioned mainly by the type of superconductive 
emittor. 

\textbf{\textit{b) The hydrogen atom}}

The density of electromagnetic energy in the hydrogen atom is
\begin{equation}
\label{eq:46a}
\varepsilon _{(H)} = {\frac{{e^{2}}}{{4\pi .\varepsilon _{o} .r_{(H)}^{2} 
}}} \cong 10^{12} \mbox{J/m}^3 
\end{equation}
Putting the hydrogen atom into the gravitational field with a higher
energy density, it would transform to a neutron. It is surely not a
coincidence that the surface of a neutron star with the mass
approaching $10^{30}$ kg and radius of about 10 km (common parameters
of neutron stars) is characterized by the gravitational energy density
close to that given by~(\ref{eq:46a}) (cf. 13). Evaluating the issue
from another angle, based on (\ref{eq13}) and~(\ref{eq:46a}) it is
possible to estimate the parameters of neutron stars.

\textbf{\textit{c) Solar corona temperature}}

The magnetic field density of a pulse magnetic field with the intensity of 
about $H \cong 7 \times 10^{3}$ A/m is identical to the absolute value of the 
Earth gravitational field density. If a pendulum was positioned in such a 
magnetic field, each pulse would cause a displacement of the pendulum. A 
mean intensity of the Sun magnetic field is about $10^{2}$ A/m, at some 
processes of magnetic field changes, however, the magnetic field intensity 
rises up to $10^{5}$ A/m. It is worth pointing out that at the intensity of 
about $4 \times 10^{4}$ A/m the magnetic field density is equal to the energy 
density of the Sun gravitational field. Interference of both the fields 
might, at certain circumstances, increase the kinetic energy of particles 
forming the solar corona by means of Podkletnov-like effect and rise its 
temperature up to the present value of $10^{6}$ K. The apparent uniformity 
of the solar corona temperature might be a consequence of periodicity of the 
magnetic effects.

\subsection*{Acknowledgements}

The financial support of the research by the Slovak Grant Agency (Projects 
VEGA/1/6106/99) is appreciated.

\bigskip

\section*{References}

\begin{description}

\item {}[1] E. Podkletnov, R. Nieminen, Physics C 203 (1992) 441

\item {}[2] E. Podkletnov, cond-mat/9701074

\item {}[3] E. Podkletnov, G. Modanese, physics/0108005

\item {}[4] V. Skalsk\'{y}, M. S\'{u}ken\'{\i}k, Astrophys. Space Sci. 178 (1991) 
169 

\item {}[5] V. Skalsk\'{y}, M. S\'{u}ken\'{\i}k, Astrophys. Space Sci. 181 (1991) 
153

\item {}[6] V. Skalsk\'{y}, M. S\'{u}ken\'{\i}k, Astrophys. Space Sci. 236 (1996) 
295

\item {}[7] J. \v{S}ima, M. S\'{u}ken\'{\i}k, gr-qc/9903090

\item {}[8] M. S\'{u}ken\'{\i}k, J. \v{S}ima, J. Vanko, gr-qc/0010061

\item {}[9] P.C. Vaidya, Proc. Indian Acad. Sci. A33 (1951) 264

\item {}[10] K.S. Virbhadra, Pramana - J. Phys. 38 (1992) 31 

\item {}[11] M S\'{u}ken\'{\i}k, J. \v{S}ima, gr-qc/0101026; Acta Phys. Pol., 
submitted 

\item {}[12] K.S. Virbhadra, Phys. Rev. D60 (1999) 104041

\end{description}

\end{document}